\documentstyle[preprint,tighten,aps]{revtex}
\begin{document}
\draft

\title{Fluid membranes in hydrodynamic flow fields:\\ 
Formalism and an
application to fluctuating \\
quasi-spherical vesicles in shear flow}

\author{ Udo Seifert\\  ~ }

\address{ Max--Planck--Institut f\"ur
Kolloid-- und Grenzfl\"achenforschung, \\
Kantstr. 55, 14513 Teltow--Seehof,
Germany.}
\maketitle

\begin{abstract} 
The dynamics of a  single fluid bilayer membrane in an
external hydrodynamic
flow field is
considered. The deterministic equation of motion for the configuration
is derived taking into account both viscous dissipation in the
surrounding liquid and local incompressibility of the membrane.
For quasi-spherical vesicles in  shear flow, thermal fluctuations
can be incorporated in a Langevin-type  equation of motion
for the deformation amplitudes.
The solution to this
equation shows an  overdamped oscillatory approach to
a stationary tanktreading shape. Inclination angle and ellipiticity of
the contour are determined as a function of excess area and shear rate.
Comparisons to numerical results and experiments are discussed.

\end{abstract}

\pacs{PACS Numbers:   47.15Gf, 68.15+e, 82.70--y}

%\narrowtext

\def\beq{\begin{equation}}
\def\ee{\end{equation}}
\def\pcite{\protect\cite}
\def\Tk{T/\kappa}
\def\kT{\kappa/T}
\def\eps{\epsilon}

\section {Introduction}

\def\lsim
{\protect \raisebox{-0.75ex}[-1.5ex]{$\;\stackrel{<}{\sim}\;$}}

\def\gsim
{\protect \raisebox{-0.75ex}[-1.5ex]{$\;\stackrel{>}{\sim}\;$}}

\def\lsimeq
{\protect \raisebox{-0.75ex}[-1.5ex]{$\;\stackrel{<}{\simeq}\;$}}

\def\gsimeq
{\protect \raisebox{-0.75ex}[-1.5ex]{$\;\stackrel{>}{\simeq}\;$}}

\def\pcite{\protect \cite}

The equilibrium behavior of fluid vesicles made up of a
single lipid bilayer membrane has been studied in great detail and is
by now  well understood \cite{z:seif96a}. 
 These equilibrium phenomena comprise the occurrence of a
multitude of vesicle
 shapes and their thermal fluctuations. If external parameters such as
temperature or osmotic conditions are changed, such a shape
may become unstable and settle into the next dynamically accessible
minimum of bending energy. The dynamics of such a shape change, however,
is a non-equilibrium process. In general, only very little is
known systematically and quantitatively about this and other
non-equilibrium phenomena of fluid vesicles.

A rough classification of non-equilibrium phenomena  
should distinguish between two types. First, there is 
relaxation into a new equilibrium after
a parameter change. This class comprises the above mentioned decay
of a metastable shape such, e.g.,  as the budding process 
\cite{kaes91,z:doeb97}. 
The best studied case of this class is  the spectacular pearling
instability of cylindrical vesicles, which develops
 upon action of optical tweezers \cite{z:bar98}.

The second class of nonequilibrium behavior refers to 
genuine non-equilibrium
states induced by external fields such as hydrodynamic flow fields.
The behavior of vesicles under such conditions has only recently been started
to be explored theoretically \cite{z:krau96,z:youh95,z:brui96} and in experiment
\cite{z:haas97}. We developed a numerical
code which allows to follow the shape evolution of vesicle in
shear flow \cite{z:krau96}. As a main result, we found that
the shape finally settles into a non-axisymmetric stationary ellipsoidal
shape around which the membrane exhibits tanktreading motion. 
Quantitative
predictions include the inclination angle and the 
tanktreading frequency as a function of reduced volume and shear rate.
This work left aside two interesting aspects.  First,  the regime
 of very small shear rates was difficult to reach within
this numerical approach because of long relaxation times. Even though the
numerical data seemed to indicate that shear is a singular perturbation,
  no clear assessment of  this regime was
possible. Secondly, fluctuations, which  would be particularly
important at small shear
rate, were not included into the numerical algorithm.

The purpose of this paper is twofold. First, we present somewhat
more explicity the underlying continuum theory for the dynamical evolution of
an incompressible fluid membrane in an external flow field thereby
generalizing previous work on dynamics of membranes in quiescent fluids
\cite{z:cai95} and within a Rouse model \cite{folt94}. 
The resulting equations are strongly non-linear and must, in general,
be solved numerically. Analytical progress, however, is possible for
so-called quasispherical vesicles 
\cite{schn84,helf86,miln87,enge85,biva87,z:yeun95,z:seif95c}
 for which deviations from a sphere
remain small throughout the parameter space. Focussing on such vesicles,
it becomes possible to include the effect of thermal fluctuations 
which is the second objective of this paper. A main effect of
thermal fluctuations in fluid membranes is that they store or ``hide''
area in the suboptical range \cite{helf84a}. 
The present work addresses the issue
of how external flow  pulls out hidden area from
thermal fluctuations. 
 Apart from its fundamental significance such a study is motivated
also by the so far only published experiments on the behavior of
vesicles in shear flow \cite{z:haas97}. In this experiment initially spherical vesicles
elongate under shear  thereby pulling out area from
thermal fluctuations. The theory developed here should be used
to analyze  such experiments.

For a somewhat broader perspective, it may be important to point out that
there exist a vast literature on the behavior of a ``soft sphere'' in
shear flow
ranging from liquid droplets \cite{rall84,z:ston94} to elastic capsules
 \cite{bart81,bart91,zhou95} or inert ellipsoids \cite{kell82,z:olla97,z:olla97a}
 modeling red-blood-cells \cite{fisc78} or synthetic
 microcapsules \cite{z:piep98}.
Many of these systems show ellipsoidal deformations with revolving or
tanktreading motion. The details of deformation,
 inclination angle and tanktreading
frequency, however, depend crucially on  both the
constitutive relation for the elasticity of the ``interface'' of the
soft sphere and the dissipative mechanisms involved. In our case, 
 the membrane is determined by bending elasticity
and local incompressibility. The main dissipation occurs in the
surrounding fluid.

How thermal fluctuations are affected by shear flow seems not to have been
touched upon in these studies of soft spheres. For the topologically
much simpler case of  on average planar membranes in a
 dilute lamellar phase,  
 the layer spacing is predicted to decrease
with increasing shear due to the suppression of 
fluctuations \cite{z:rama92,z:brui92}.

This paper consists of five sections, of which this is the first. Section II
discusses the physical model underlying this work
 and derives the fundamental equation of motion for
an incompressible membrane in external flow. In Section III, we specialize to
quasi-spherical vesicles in linear flow. In Section IV,
we discuss in detail shear flow for which we solve the equations 
explicitly and compare
to available numerical and experimental work.
A summarizing discussion is presented in Section V.

\def\r{{\bf r}}
\def\vi{{\bf v}^{\rm ind}}
\def\vu{{\bf v}^{\infty}}
\def\Vu{{\bf V}^{\infty}}
\def\Vi{{\bf V}^{\rm ind}}
\def\Y{{\cal Y}_{l,m}(\theta,\phi)}
\def\Xi{X^{{\rm ind}}}
\def\Yi{Y^{{\rm ind}}}
\def\Xu{X^{\infty}}
\def\Yu{Y^{\infty}}

\def\R{{\bf R}}
\def\ss{(s_1,s_2)}

\def\u{u_{l,m}}
\def\a{|v_{l,m}|^2}
\def \Z{Z^\infty}
\def\k{\kappa}
\def\D{\Delta}
\def\p{\partial}
\def\dep{\delta p}\def\dl{\delta L}

\section{Formalism}
\subsection{Physical model}
The starting point for the dynamics of a fluid membrane
 is the bending energy \cite{helf73}
\beq
F_\kappa \equiv {\kappa\over 2} \int dA (2H-C_0)^2 .
\label{eq:Fk}
\ee Here, $\kappa$ is the bending rigidity, 
$H$ is the local mean curvature and $C_0$ a spontaneous curvature
in the case of intrinsically asymmetric monolayers or an asymmetric
liquid environment. We neglect 
Gaussian curvature energy which would enter only
through boundary values, since we focus on the dynamics of
a closed membrane.
For simplicity, we also neglect the fact that for closed membranes
a second energy term becomes relevant which takes into account that
the bilayer consists of two tightly coupled monolayers which do not 
exchange molecules \cite{miao94}. 

The bending  energy is  the driving force for dynamical changes. Dynamics in
the micron world of vesicles is overdamped, i.e. inertial effects
can safely be ignored as can easily be checked a posteriori by
calculating the corresponding Reynolds number \cite{happ73}. Dissipation takes place
both in the surrounding liquid and in the membrane, in principle.
For giant vesicles of 
micron size, the dominant dissipation  is viscous dissipation
in the embedding fluid \cite{kram71,broc75a}.  Therefore, a full treatment
of the hydrodynamics of this fluid is mandatory for a 
faithful description of the dynamics of membranes. 

Dissipation in the membrane
can be classified into three phenomena: Drag between the
two monolayers \cite{evan91}, shear viscocity within each layer and permeation through
the membrane. Calculation of the relaxation spectra of bending fluctuations
involving the
first two mechanisms show that on scales of microns and larger, hydrodynamic
dissipation is dominant \cite{seif93a}. In the submicron range, 
friction between the layers becomes relevant. On even smaller scales of
several tenths of nanometers, shear viscosity within each layer
should be included.  Finally, permeation through the membrane seems to
be irrelevant on all length-scales with the possible exception of
membranes in the vicinity of a substrate \cite{z:pros98}. 

Based on this hierarchy of dissipative mechanisms, we will 
include only hydrodynamic dissipation and
 model the membrane  as impermeable for liquid.
As a consequence, the normal velocity of the fluid at the 
membrane pushes along the membrane and leads to a shape change.
Tangential motion of the  fluid along the membrane 
induces lipid flow within the membrane since we employ non-slip
boundary conditions between fluid and membrane. 
One could allow for some slip  with a
phenomenological friction coefficient. In the absence of any
evidence for such a phenomenon, however, we choose
 for simplicity the no-slip condition used so far. Finally, we require that the membrane remains
locally incompressible.

\subsection {Elementary differential geometry of the membrane}

The instantaneous membrane configuration 
$\R\ss$, parametrized by internal coordinates $\ss$,
 is embedded in the three dimensional space. This space
will be parametrized by Cartesian $(x,y,z)$ or spherical $(r,\theta,\phi$)
 coordinates as
$\r=x_\alpha{\bf e}_\alpha = r {\bf e}_r$ where
$\alpha =x,y,z$. 
Summation over double indices is implied throughout the paper. 
 There are two 
tangential vectors 
\begin{equation}
{\bf R}_i \equiv   \partial_i {\bf R}(s_1,s_2)
 \quad \hbox{\rm for} \quad i=s_1, s_2  ,
\end{equation}
from which one obtains the metric tensor
\beq
g_{ij}\equiv {\bf R}_i \cdot {\bf R}_j  .
\ee
Its determinant, 
$g\equiv \det(g_{ij})$, 
yields the area element
\beq
dA=\sqrt g\; ds_1 ds_2 \label{eq:dA} .
\ee
The normal vector, ${\bf n}(s_1,s_2)$,  is given by 
\begin{equation}
{\bf n} = \frac{{\bf R}_1 \times {\bf R}_2}
{|{\bf R}_1 \times {\bf R}_2|}  . 
\end{equation}
 Finally, the mean and the Gaussian curvature follow from the
curvature tensor  
\begin{equation}
h_{ij} \equiv  (\partial_i \partial_j{\bf R})\cdot {\bf n}\label{eq:hij}
\end{equation}
as 
\beq
H\equiv  {1\over 2}  \ h^i_{\ i} 
\label{eq:H}
\ee
and
\beq
K\equiv det (h^i_{\ j}) ,
\label{eq:K}
\ee 
where
$h^i_j\equiv g^{ik}h_{kj}$ and $g^{ij}$ are the matrix
elements of the matrix inverse of $(g_{ij})$.
Following the convention used in differential geometry,
a sphere with the usual spherical coordinates ($s_1=\theta, s_2=\phi)$
has $H<0$. 
A membrane configuration
has bending energy $F_\kappa$ given in (\ref{eq:Fk}). 

In order to ensure local incompressibility of the membrane, we will
need a local Lagrange multiplier $\Sigma\ss$ which we call local
surface tension. It will be determined self-consistently below.
The total ``energy'' thus becomes
\beq
F\equiv F_\kappa + \int ds_1ds_2\sqrt{g}\  \Sigma\ss .
\label{eq:Fe}
\ee

\subsection {Equation of motion}

Any non-equilibrium  membrane configuration exerts a local three dimensional
force density
\beq
{\bf f}(\r) \equiv - \int ds_1ds_2\sqrt{g} \left({1 \over \sqrt{g}}{\delta
F\over \delta \R}\right)\delta(\r -\R\ss)
\label{eq:ff}
\ee
onto the surrounding fluid.

The variational derivative entering the force density
reads explicitly
\beq
\left({1 \over \sqrt{g}}{\delta
F\over \delta \R}\right)
= \left(-2\Sigma H +\kappa [(2H+C_0)(2H^2-2K-C_0H)+ 2\Delta H]
\right){\bf n}
 -g^{ij}\R_i\p_j\Sigma .
\label{eq:se}
 \ee
Here, 
\beq
\Delta\equiv (1/\sqrt g) \partial_i(g^{ij}\sqrt g \partial_j)
=g^{ij}\p_i\p_j .
\label{eq:LB}
\ee
 is the 
 Laplace--Beltrami operator on the surface.
The normal part of (\ref{eq:se})
 is well-known from the stationarity condition 
of membrane configurations \cite{ouya89}. The tangential part arises from inhomogeneities
in the surface tension which will be needed to ensure local 
incompressibility of the
induced flow. This approach differs from the one used in Ref. \cite{z:cai95}
where a finite compressibility was introduced for renormalization purposes.

The surrounding liquid is incompressible
\beq {\bf \nabla} {\bf v} =0
\ee 
and 
obeys the Stokes equations
\beq
{\bf \nabla} p -\eta {\bf \nabla}^2 {\bf v} = {\bf f}(\r) . 
\ee
A special
solution of the inhomogeneous Stokes equation for the
 velocity field reads
\beq
\vi(\r)= \int d\r' {\cal O}(\r,\r') {\bf f}(\r') ,
\label{eq:vii}
\ee
where the Oseen tensor ${\cal O}(\r,\r')$ has Cartesian matrix elements \cite{doi86}
\beq
{\cal O}_{\alpha\beta}(\r,\r')\equiv 
{1\over 8 \pi \eta |\r-\r'|}\left [\delta_{\alpha\beta}+{(r_\alpha
-r'_\alpha)(r_\beta-r'_\beta)
\over |\r-\r'|^2}\right ] .
\ee
Thus, the hydrodynamics mediates a long-range interaction 
($\sim
1/ |\r-\r'|$)  through
the velocity field.
To this velocity field $\vi(\r)$ induced by the presence of the membrane, we
must add the externally applied flow field $\vu(\r)$ to obtain the
total velocity field
\beq
{\bf v}(\r) = \vu(\r) +\vi(\r) .
\ee
Such a simple superposition becomes possible since the Stokes equation
 is linear due to the absence of the convective term. 
The value of the total velocity field at  the position of the membrane yields
the equation of motion for the membrane configuration
\beq
\p_t \R\ss =  \vu(\R\ss) +\vi(\R\ss)  ,
\label{eq:pt}
\ee
since we employ no-slip boundary conditions at the
membrane. This deterministic equation of motion includes  both normal
motion that signifies a shape change of the membrane and
tangential motion that corresponds to lipid flow
within the membrane. The so far
unknown local tension $\Sigma\ss$ is determined by
requiring that the membrane flow induced by this equation obeys
local incompressibility. Thus, we must demand
\beq
\p_t\sqrt{g} = 0 ,
\ee
which implies explicitly
\beq
g^{ij}\R_i\p_j(\p_t \R) =0 .
\label{eq:gij} \ee
Upon insertion of the equation of motion (\ref{eq:pt})
with (\ref{eq:ff}) and (\ref{eq:vii}), this condition becomes a 
 partial differential equation for the unknown tension $\Sigma\ss$. 

Equations (\ref{eq:pt}) and (\ref{eq:gij}) yield a
  deterministic evolution
equation for a membrane configuration under the action of bending energy
and hydrodynamics. For any given initial membrane configuration, the
solution to these equations will run into the next
dynamically accessible local minimum of bending energy. In general, these
equations must be solved numerically. For vesicles in shear flow, this
has been achieved recently \cite{z:krau96}.

\section{Quasi-spherical vesicle in flow}

In this section, we apply the general formalism to quasi-spherical
 vesicles 
\cite{schn84,helf86,miln87,enge85,biva87,z:yeun95,z:seif95c}
for which analytical progress becomes possible.

\subsection{Expansion around the sphere}

A quasispherical vesicle is characterized by its volume 
\beq
 V\equiv {4 \pi\over 3} R^3,
\label{eq:Vcon}
\ee
which  defines $R$, and its fixed  area
\begin{equation} A\equiv (4\pi +\Delta) R^2, 
\label{eq:A}
\end{equation}
which defines the (dimensionless) excess area $\Delta$. 

The instantaneous vesicle shape 
$\R(\theta,\phi) \equiv R(\theta,\phi){\bf e}_r$ 
can be parametrized by  spherical harmonics
\begin{equation}
R(\theta,\phi)=R\left(1+\sum_{l\geq 0}^{l_{max}}\sum_{m=-l}^{l}
 \u {\cal Y}_{lm}(\theta,\phi) \right)  ,
\end{equation}
where $|m|\leq l$ and  $u_{l,-m}=(-1)^m\u^*$.
The upper cutoff $\l_{max}$ is of order $R/d$
where $d$ is the membrane thickness. For a vesicle with
$R\simeq 10 -50 \mu m,\  \l_{max} $ is of order $10^4$.
Since spontaneous curvature is irrelevant for 
the shape behaviour of quasi-spherical vesicles \cite{z:seif95c}, we set for the
rest of the paper $C_0=0$. 
Expanding the geometrical quantities as well as the bending energy
around a sphere, one has \cite{helf86,miln87,ouya89}
\beq
F_\kappa= 8\pi + 
  {1\over 2} \sum_{l\geq 0}^{l_{max}}\sum_{m=-l}^{l}
 |\u|^2(l+2)(l+1)l(l-1) + O(u_{l,m}^3) ,
\label{eq:Gfluc}
\ee
\beq
A= R^2 \left(4\pi\left(1+{u_{0,0}\over\sqrt{4\pi}}\right)^2 +  \sum_{l\geq 1}^{l_{max}}\sum_{m=-l}^{l}
|\u|^2(1+l(l+1)/2) + O(u_{l,m}^3) \right)  ,
\label{eq:Afluc}
\ee
and \beq
V= R^3 \left({4\pi\over 3} 
\left(1+{u_{0,0}\over\sqrt{4\pi}}\right)^3 
+  \sum_{l\geq 1}^{l_{max}}\sum_{m=-l}^{l}
|\u|^2 + O(u_{l,m}^3) \right)  .
\label{eq:Mfluc}
\ee
The volume constraint (\ref{eq:Vcon}) fixes the amplitude $u_{0,0}$ as a function of the
other amplitudes
 \beq
u_{0,0}=-\sum_{l\geq 1}^{l_{max}}\sum_{m=-l}^{l}|\u|^2/\sqrt{4\pi}  ,
\label{eq:vc}
\ee
where we truncate from now on the cubic terms. If this value is inserted into (\ref{eq:Afluc}), 
the area constraint (\ref{eq:A}) becomes 
\beq
\sum_{l\geq 1}^{l_{max}}\sum_{m=-l}^{l}
|\u|^2{(l+2)(l-1)\over 2}=\Delta
\label{eq:D0}
\ee
 Since the $(l=1)$-modes
correspond to translations, which have to be omitted, from now on
 all sums start at $l=2$. We will abbreviate
\beq
\sum_{l,m} \equiv \sum_{l\geq 2}^{l_{max}}\sum_{m=-l}^l  .
\ee
We now add the global area constraint (\ref{eq:D0})
 with a Lagrangean multiplier 
\beq
\Sigma
\equiv \kappa \sigma/R^2 \ee
to the quadratic part of the curvature energy. This leads to a quadratic
expression for the total ``energy'' (\ref{eq:Fe})
\beq
F={\kappa\over 2} \sum_{l,m} E_l |\u|^2 ,
\ee
with 
\beq
E_l=(l+2)(l-1)[l(l+1)+ \sigma ] .
\label{eq:El}
\ee

 The instantaneous state of  a vesicle must be characterized
not only by the set of its deformations ${\u}$ but also by its
instantaneous surface tension 
\beq
\Sigma(\theta,\phi)\equiv {\kappa \over R^2} 
(\sigma +  \sum_{l,m} \sigma_{l,m} \Y ).
\ee The homogeneous value $\sigma=\sigma_{0,0}$, which will be called
effective tension, has already been included into the energy (\ref{eq:El}).

\subsection{Velocity field}

The notion of a quasi-spherical vesicle implies that the
deviations $u(\theta,\phi)$ from the spherical shape are small. Then
 stresses caused by bending moments
and an inhomogeneous surface tension can be assumed to act on the sphere
rather than on the deformed vesicles's surface. Likewise, all velocity fields
will be evaluated and matched at the sphere. We retain this procedure for
arbitrary flow strength even though it is strictly valid only for small
external fields. As will be discussed in Sect. IV.D below, while not
modifying scaling laws, this approximation
causes an error in numerical prefactors at large shear rates of at most 30 \%.

The instantaneous stress distribution caused 
by both the deformation and the inhomogeneous
 surface tension induces an
irrotational velocity field $\vi(\r)$ which we determine in the Appendix
adapting the classical Lamb solution \cite{happ73}. 
 For this purpose, it is convenient to
define  another
three-dimensional velocity field 
\beq
\Vi(\r) \equiv \vi(R,\theta,\phi)
\ee
which extends the velocity field on the sphere corresponding to the
vesicle formally to all space. $\Vi(\r)$ is $r$-independent.
 This velocity
field can then be characterized by its normal component
\beq
\Xi \equiv \sum_{l,m}\Xi_{l,m}\Y\equiv \Vi(R,\theta,\phi) {\bf e}_r
\ee
and the negative of its divergence
\beq
\Yi\equiv \sum_{l,m}\Yi_{l,m}\Y\equiv  -R\  {\bf \nabla} \Vi  ,
\ee where $\Y$ are the usual spherical harmonics.
Although  the full three-dimensional velocity field 
$\vi(\r)$ is
divergenceless, the derived quantity $\Vi(\r)$, in general, is not.

Normal and tangential stress balances on the vesicle surface determine
    $\vi(\r)$ as a function of the displacement $u(\theta,\phi)$ and the
surface tension $\sigma(\theta,\phi)$. As shown in the Appendix, 
we can determine the linear relationship between the expansion
coefficients of the induced velocity on the
sphere $\{\Xi_{l,m},\Yi_{l,m}\}$ and the sources of this field 
 $\{{\u, \sigma_{l,m}}\}$.
Replacing $\sigma_{l,m}$ by $\Xi_{l,m}$ and $\Yi_{l,m}$, one finally gets
 \beq
\Xi_{l,m} = - (\kappa/\eta R^2)\ \Gamma_l E_l \u - B_l \Yi_{l,m}  ,
\label{eq:Xi}
\ee 
where \beq
\Gamma_l \equiv {l(l+1)\over 4l^3+6l^2-1}
\ee
and
\beq
B_l \equiv {2l+1\over  4l^3+6l^2-1}  .
\label{eq:Bl}
\ee

It is convenient to characterize the  external flow field $\vu(\r)$
in a similar manner. We therefore define
\beq
\Vu(\r) \equiv \vu(R,\theta,\phi) ,
\ee
 its normal component
\beq
\Xu\equiv \sum_{l,m}\Xu_{l,m}\Y \equiv \Vu(R,\theta,\phi) {\bf e}_r
\ee
and its divergence
\beq
\Yu\equiv \sum_{l,m}\Yu_{l,m}\Y\equiv - {\bf \nabla} \Vu  .
\ee

The equation of motion (\ref{eq:pt}) for the 
quasi-spherical vesicle in this external flow thus
becomes
\beq
\p_t \R (\theta, \phi) = \Vu(\theta, \phi) + \Vi(\theta, \phi) .
\label{eq:motion}
\ee
The local incompressiblility condition (\ref{eq:gij}) 
becomes after little 
algebra
\beq
{\bf \nabla}(\Vu + \Vi) =0 ,
\ee
or
\beq
\Yi=-\Yu .
\label{eq:Yi}
\ee
 Physically, the incompressibility condition fixes the local
surface tension which we have already eliminated in favor of $\Yi$.

 We will now
set up the equation of motion for the normal deviation,
$u(\theta,\phi)$,  from the
perfect sphere,  where $\theta$ and $\phi$ are
fixed coordinates in the lab frame.
It is important to realize that (\ref{eq:motion}) leads to a tangential
motion of the material point labeled by $(\theta,\phi)$. Therefore it is not sufficent to
project (\ref{eq:motion}) onto ${\bf e}_r$
in order to get the normal shape change in the lab frame. We also  have
to include an advection
term as for planar membranes \cite{z:rama92,z:brui92}. The equation of motion thus becomes
\beq
\p_t u(\theta,\phi) = - (\Vu + \Vi) {\bf \nabla} u(\theta,\phi) + 
(\Xu +\Xi)/R  .
\label{eq:u}
\ee
Note that the advection term can be written in terms of three-dimensional
vectors because $u$ carries no $r$ dependence. In general, the advection
term will couple the different modes. Further progress becomes possible
for the quite wide class of linear external flow.

\subsection {Langevin equation for linear flow}

Linear external flow is characterized by the form
\beq
\vu(\r) = {\cal G} \r
\ee
where ${\cal G}$ is a traceless $\r$-independent matrix. 
It can be decomposed according to
\beq
\vu(\r) = {\cal G}_s \r + {\bf \Omega} \times \r  ,
\ee
where ${\bf\Omega}$ denotes magnitude and direction of the rotational part
of the flow and ${\cal G}_s$ is a symmetric traceless matrix
describing the strain or irrotational component
 of the flow. The strain component  of the external
flow is compensated by the corresponding induced flow because of the
incompressibility constraint. 
 The advection term thus involves only the rotational component of
the flow and reads 
\beq
- ({\bf\Omega} \times \r){\bf \nabla} u(\theta,\phi)= -i{\bf\Omega}
 ({\bf L}/\hbar)u(\theta,\phi) ,
\ee
where ${\bf L}\equiv \r \times (-i\hbar {\bf \nabla})$ 
is the angular momentum operator. Choosing co-ordinates
such that ${\bf \Omega} = \Omega {\bf e}_z$, we finally obtain 
from (\ref{eq:u}) 
the equation of motion 
\beq
\p_t \u =  - i \Omega m \ \u  - (\kappa/\eta R^3)\Gamma_l E_l \ \u
+ (\Xu_{l,m} + B_l \Yu_{l,m})/R
 +\zeta_{l,m}
\label{eq:u2}
\ee
after expanding
in 
 spherical harmonics  and using (\ref{eq:Xi}) and (\ref{eq:Yi})

In order to determine 
the correlations of the thermal noise $\zeta_{l,m}$ we apply
this equation of motion to equilibrium \cite{schn84,helf86,miln87}. Then, there is no external flow,
$\Omega = \Xu_{l,m} = \Yu_{l,m} = 0$. In the long time limit,
the dynamical equal time correlations 
calculated from (\ref{eq:u2}) will reproduce  the static correlations only
if we choose
\beq
<\zeta_{l,m}(t)\zeta_{l',m'}(t')> = 2 T (R^3/\eta)\Gamma_l 
(-1)^m \delta_{l,l'} \delta_{m,-m'} \delta (t-t').
\label{eq:noise}
\ee
Here, $T$ is the temperature and Boltzmann's constant is 
set to 1 throughout.

We keep the noise correlations (\ref{eq:noise})
 in the presence of external flow. While 
this is certainly correct for small flow 
it is not clear whether and how
strong flow modifies these correlations. 
Note in passing that we have refrained from
adding noise  to the general equation of motion
(\protect\ref{eq:pt}), since 
 the appropriate correlations for 
these forces are not clear for the same reason. Naively,
one would expect them to exhibit long-range spatial correlations given
by the Oseen tensor for small flow. Moreover, even in equilibrium,
 there are subtleties associated both with
the incompressibility constraint and measure factors \cite{z:cai95}.

The stationary value $\bar \u$ of the amplitude $\u(t)$ follow
from (\ref{eq:u2}) as
\beq
\bar \u = \left({\eta R^2\over \kappa}\right)\left(
{\Xu_{l,m} + B_l \Yu_{l,m} \over \Gamma_l E_l + i \tilde
\Omega m}\right) ,
\label{eq:baru}
\ee
where 
\beq
\tilde \Omega\equiv \Omega \eta R^3/\kappa  .
\ee
For linear flow, $\bar \u \not = 0$ only for $l=2$.

The deviations from this stationary value,  
\beq
\eps_{l,m}(t) \equiv \u(t) -\bar \u ,   
\ee
obey the equation of motion
\beq
\p_t\  \eps_{l,m} =    - i \Omega m \ \eps_{l,m} 
 - (\kappa/\eta R^3)\Gamma_l E_l\  \eps_{l,m} +\zeta_{l,m} .\ee   
This simple equation is easily solved as
\beq
\eps_{l,m}(t) = \exp[( - (\kappa/\eta R^3)\Gamma_l E_l - i 
\Omega m)t]
(\int_o^t d\tau \exp[ ((\kappa/\eta R^3)\Gamma_l E_l +i
\Omega m)\tau]
\zeta_{l,m}(\tau) + \eps_{l,m}(0)) .
\ee     
Starting from an initial value $\eps_{l,m}(0)$, the approach to
the stationary value (\ref{eq:baru}) thus happens via
 damped oscillations.
                                               
Using the noise correlations (\ref{eq:noise}), the dynamical correlation function
in the long time limit 
become
\beq
\lim_{t\to\infty} <\eps_{l,m}(t)\eps_{l,-m}(t+\Delta t)> = (-1)^m 
\exp[(-(\kappa/\eta R^3)\Gamma_l E_l + i m \Omega)\Delta t]
{T\over \kappa E_l}  
\ee
with the stationary correlations
\beq
<\eps_{l,m}(t)\eps_{l,-m}(t)> = (-1)^m 
{T\over \kappa E_l}  .
\ee

\subsection {Area constraint and effective tension}

As a last step, we have to eliminate the  yet
unknown Lagrange multiplier or effective tension 
$\sigma$ in favor of physical quantities using the area constraint 
(\ref{eq:D0}).
Excess area is stored both in the systematic stationary amplitudes
$ \bar \u$ as well as in fluctuations $\eps_{l,m}$. For the
systematic part, we get 
\beq
\bar \Delta \equiv \sum_{l,m}{(l+2)(l-1)\over 2}
|\bar \u|^2 = 2   
 \left({\eta R^2\over \kappa}\right)^2 \sum_{m=-2}^2 
\left({(\Xu_{2,m} + B_2 \Yu_{2,m})^2
 \over \Gamma_2^2 E_2^2 + \tilde \Omega^2 m^2 }\right) .
\label{eq:bardelta}
\ee
 The fluctuation 
contribution $\Delta_{l,m}$ of the mode  $\epsilon_{l,m}$ to the excess area  is
\beq
\Delta_{l,m}\equiv 
 {(l+2)(l-1)\over 2} |\eps_{l,m}|^2 = 
{(l+2)(l-1)\over 2}{T\over \kappa E_l}.
\ee
The total area constraint becomes
\beq
\Delta = 2 
 \left({\eta R^2\over \kappa}\right)^2
\sum_{m=-2}^2
\left({(\Xu_{2,m} + B_2 \Yu_{2,m})^2
 \over \Gamma_2^2 E_2^2(\sigma) +  \tilde \Omega^2 m^2}\right) +
{T\over 2 \kappa }\sum_{l,m}
{ (l+2)(l-1)\over  E_l(\sigma)} ,
\ee
where we made the $\sigma$-dependence of $E_l$ explicit. 
This master equation determines the unknown effective tension
 $\sigma$  as a function 
of the physical parameters $T/\kappa$ and those characterizing 
the external flow. In general, it must be solved numerically.

\section {Shear flow}
In this section, we specialize the general results of the previous section to 
the experimentally important case of simple shear flow with shear rate 
$\dot \gamma$.

\subsection {Flow parameters}
We choose the coordinate system such that the externally
imposed simple shear flow reads
\beq
\vu(\r) = \dot \gamma y {\bf e}_x = \dot \gamma [ (y/2){\bf e}_x
+(x/2) {\bf e}_y)]  -(\dot \gamma/2) {\bf e}_z \times \r .
\ee
We have separated the rotational component and identify 
\beq
\Omega = -\dot \gamma/2 {\ \ \ \rm and \ \ \ } \tilde \Omega =  - \chi/2  ,
\ee
where
\beq
\chi \equiv \dot \gamma \eta R^3/\kappa 
\ee
is the dimensional shear rate. 
The elongational first part  of the shear flow can be written as
\begin{eqnarray}
\dot \gamma [ (y/2){\bf e}_x
+(x/2) {\bf e}_y)] =&& \dot \gamma r (\sin^2\theta \sin \phi
\cos\phi {\bf e}_r + \cr &&
(1/2) \sin \theta 
(\cos^2\phi-\sin^2\phi){\bf e}_\phi
+ \sin\theta\cos\theta\sin\phi\cos\phi {\bf e}_\theta) .
\end{eqnarray}
One thus has
\beq
\Xu = \Yu = \dot \gamma R \sin^2\theta \sin\phi\cos\phi
\ee
and identifies the expansion coefficients in spherical harmonics as
\beq
\Xu_{2,\pm2}=\Yu_{2,\pm2}=\mp i \dot \gamma R (2\pi/15)^{1/2}  
\ee
and any other $\Xu_{l,m} = \Yu_{l,m} = 0$.
Note as an aside that $\Xu = \Yu $ holds for any linear flow.

\subsection {Effective tension}

 We must eliminate the effective tension $\sigma$
 in favor of the area constraint. 
The external shear flow implies the non-zero stationary amplitudes 
(\ref{eq:baru}) 
\beq
\bar u_{2,\pm2} = 
%\left({\eta R^2\over \kappa}\right)
%\left({\mp i \dot \gamma (12/11) (2\pi/15)^{1/2}
%\over
%\Gamma_2 E_2 \mp i \dot \gamma}\right) 
{\mp i \chi 
 (12/11) (2\pi/15)^{1/2} \over \Gamma_2E_2\mp i \chi}
  ,
\ee
where we used $B_2 = 1/11$.
According to (\ref{eq:bardelta}) the systematic part of excess area stored in the modes $\bar u_{2,\pm2}$
contributes
\beq
\bar \Delta = 2(|\bar u_{2,2}|^2 + |\bar u_{2,-2}|^2) = a_2 {\chi^2\over 
\Gamma_2^2E_2^2 + \chi^2}
\label{eq:Db}
\ee
where
\beq
a_2 \equiv 4 (12/11)^2{2\pi/15}\simeq 1.994 .
\ee

The area stored in the fluctuations becomes
\beq
{T\over 2 \kappa} \sum_{l,m} {(l+2)(l-1)\over E_l} = {T\over 2\kappa}
 \sum_{l\geq 2}
{2l+1\over l(l+1) + \sigma} = {T\over 2 \kappa}{5\over 6 + \sigma} + 
{T\over 2 \kappa} \sum_{l\geq 3} {2l+1\over l(l+1) + \sigma}.
\ee
Replacing the last sum by an integral,
we obtain the total area constraint in the form
\beq
a_2{\chi^2\over 
\Gamma_2^2E_2^2 + \chi^2} +  {T\over 2 \kappa}\left({5\over 6 + \sigma}\right)
+{T\over 2 \kappa}\ln\left({l_{max}^2+\sigma\over 12+\sigma}\right) = \Delta .
\label{eq:delta}
\ee
This equation can easily be solved numerically for $\sigma = \sigma(\chi,
T/\kappa,l_{max})$. However, it is more
 instructive to discuss limiting cases and the general behavior analytically.

For vanishing shear rate, i.e., in
 equilibrium, the area constraint implies the three
 regimes \cite{z:seif95c}):

(i){\it Tense regime:}  
For $\Delta << T/2\kappa$, one obtains from (\ref{eq:delta})
\beq
\sigma \approx {T\over 2 \kappa \Delta} l_{max}^2  
\label{eq:sigma11} .
\ee
In this regime,  all $N$
modes share the available excess area.

(ii) {\it Entropic regime:} 
For  $T/2\kappa << \Delta << (T/\kappa) \ln l_{max} $, the tension
depends exponentially
on the excess area \cite{helf84a}
\beq
\sigma\approx {l^2_{max}} e^{-2\kappa\Delta/T} .  
\label{eq:sigma3}
 \end{equation} 

(iii) {\it Prolate regime:} For $(T/\kappa) \ln l_{max}^2 << \Delta \lsim 1$,
most of the excess area is stored in the $(l=2)$-modes. The tension
approaches the limiting value $-6$ \cite{miln87,z:seif95c}:
\beq
 \sigma\approx -6 + {5\over 2}{T\over \kappa\Delta}.
\label{eq:sigma1}
\ee
A vesicles that is relaxed with respect to its volume constraint
also belongs to this regime with $\sigma =0$.

For large shear rate, i.e., formally
$\chi\to\infty$, 
the whole
excess area is stored in the $\bar u_{2,\pm2}$ deformation.
From
\beq
\bar \Delta = 2(|\bar u_{2,2}|^2 + |\bar u_{2,-2}|^2) = a_2 {\chi^2\over 
\Gamma_2^2E_2^2 + \chi^2} = \Delta ,
\ee
one obtains
\beq
\Gamma_2E_2 = \chi\left({a_2-\Delta\over \Delta}\right)^{1/2} 
\label{eq:gammasat}
\ee
and hence, with $E_2\approx 4 \sigma$ and $\Gamma_2=6/55$, 
 the limiting behavior
\beq
\sigma\approx (55/24)a_2^{1/2}\chi/\Delta^{1/2}\simeq 3.24 \chi/\Delta^{1/2}.
\label{eq:sigmaasy}
\ee Thus, for large shear rate, the tension increases linearly
with shear rate with a prefactor that depends on the excess area.

The cross-over between the equilibrium tension and this shear rate dominated
saturation regime happens at
 a crossover  $\chi_c$. The scaling
behavior of  $\chi_c$ can be obtained by defining  $\chi_c$ as the
value for which half of the excess area is stored in the systematic
contribution $\bar \Delta$. This leads to the expression
 \beq
\chi_c \sim   \left\{\matrix{\displaystyle
{l_{max}^2 T/\kappa\Delta}, & {\rm \\ tense,} \cr
\noalign{\medskip}
\displaystyle l_{max}^2e^{-2\kappa\Delta /T}, &  {\rm \\ entropic,} \cr
\noalign{\medskip}
\displaystyle T/\kappa\Delta, &  {\rm \\ prolate} .\cr}
\right.\label{eq:chic}
\ee

\subsection {Shape parameters}
Knowing the value of the effective
tension as a function of the excess area, we can determine the 
characteristic parameters of the shear rate dependent shape of a quasi-spherical
vesicle.

The deviation of the mean contour from a circle
in the plane $z=0$, i.e., $\theta = \pi/2$,  becomes
\beq
u(\pi/2,\phi) = 2 {\cal R}e \left\{u_{2,2} {\cal Y}_{2,2}(\pi/2,\phi)
\right\}
={6\over 11}
{ \chi \over (\Gamma_2^2E_2^2 + \chi^2)^{1/2}}\cos2(\phi-\phi_0)  ,
\ee
with the mean incination angle \beq
\phi_0 = {1\over 2} \arctan {\Gamma_2E_2\over \chi} .
\ee
Thus, the mean incliniation angle is $\phi_0=\pi/4$ for small shear flow.
For large shear rate,  the limiting behavior
\beq
\phi_0\approx{1\over 2} \arctan\left({a_2-\Delta\over \Delta}\right)^{1/2} 
\approx\pi/4 -{\Delta^{1/2} \over 2 a_2^{1/2}} 
\ee
follows from (\ref{eq:gammasat}).
Thus, the more excess area is available, the smaller the inclination
angle. The cross-over between the two limiting cases happens
at the crossover shear rate $\chi_c$ given in (\ref{eq:chic}).

Experimentally, the ellipticity of such a contour is often measured by
the deformation parameter
\beq
D\equiv {L-B\over L+B} = 
{6\over 11}{\chi \over (\Gamma_2^2E_2^2 + \chi^2)^{1/2}},
\label{eq:D}
\ee
where $2L$ and $2B$ are the long and short axes, resp., of the contour.
For small shear rate $\chi$, the equilibrium scaling of the
tension (\ref{eq:sigma11}) - (\ref{eq:sigma1}) in the three regimes 
implies  the following linear behavior of the deformation parameter
$D$ on 
$\chi$ as
\beq
D\approx {6\over 11}{\chi \over \Gamma_2E_2}=
{6\over 11}{\chi \over (24/55)(6+\sigma)} \approx 
{5\over 4}\chi \cdot
 \left\{\matrix{\displaystyle
{2\kappa \Delta\over T l_{max}^2}, & {\rm \\ tense ,} \cr
\noalign{\medskip}
\displaystyle e^{2\kappa\Delta /T}/l_{max}^2, &  {\rm \\ entropic,} \cr
\noalign{\medskip}
\displaystyle(2\kappa\Delta/5T), &  {\rm \\ prolate} .\cr}
\right.
\ee

With increasing shear rate, deviations from this linear behavior
become important at the crossover  $\chi_c$ (\ref{eq:chic}). 
For $\chi>>\chi_c$,
the deformation saturates at the value 
\beq
D\approx \sqrt {15 \Delta/32\pi},
\ee
obtained by  combining (\ref{eq:D}) with (\ref{eq:Db}).

\subsection {Comparison to numerical work}

For large shear rate, fluctuations become irrelevant. In this regime, we can
make contact with previous numerical work where the continuum equation of
motions (\ref{eq:se}) and (\ref{eq:gij}) were solved on a triangulated
 mesh \cite{z:krau96}. 
The two quantities that can be compared within the two approaches
are the dependence of
  the inclination angle
and the effective tension on the excess area. In the previous work,
 reduced
volume
\begin{equation}
v \equiv \frac{V}{(4\pi /3) (A/4\pi)^{3/2}} = (1+\Delta/4\pi)^{-3/2} 
\label{eq:vred}
\end{equation}
was used instead of the excess area. This definition implies
\beq
\Delta = 4\pi(v^{-2/3}-1)  .
\ee
Hence the present theory predicts for large shear rate and $1-v<<1$
the limiting behavior 
\beq
\phi_0
\approx{\pi\over 4} -\left({2\pi (1-v)\over 3 a_2}\right)^{1/2}
\simeq  {\pi\over 4} - 1.025 (1-v)^{1/2}.
\ee This square-root scaling  is in quantitative agreement with the 
 previous numerical solution \cite{z:krau96}. The numerical
prefactor within the present approach is about 10\% larger than
the one extracted from  Fig. 2 of Ref. \cite{z:krau96}.
Likewise, the scaling of the  effective tension (\ref{eq:sigmaasy})
agrees with the numerical data \cite{z:krau96a} with a difference in
prefactor of about 30\%. 

The origin of the numerical difference in prefactors presumably lies
 in neglecting higher order
 couplings between deformation and external flow in the equation of motion
(\ref{eq:u2}). These effects would lead to non-diagonal and non-linear terms
in the right hand side of (\ref{eq:u2}) which would spoil the solvability. It is therefore
gratifying to know that at least for the inclination angle of the tanktreading
state, the present treatment is  quite reliable for $v\geq 0.8$ at any
shear rate.

\subsection{Comparison to experiment}

There seems to be only one published experiment on quasi-spherical
vesicles in shear flow \cite{z:haas97}. In this work, two different types of 
behavior have been reported. (i) Initially spherical vesicles rotate
at low shear rate and deform at higher shear rates into a tanktreading
state with an inclination angle close to $\pi/4$. (ii) Initially non-spherical
vesicles undergo a periodic flipping motion at small shear rate. This
flipping motion shows oscillations at larger shear rate. Only for
sufficiently high shear rate, stationary behavior with tanktreading is
observed.

The first class of behavior fits  well into the theoretical description
developed here. The experimentally found $D(\dot \gamma)$ curve shows linear
behavior at small $\dot \gamma$ and seems to saturate at larger shear rate.
de Haas et al  \cite{z:haas97}
analyzed this curve using a theoretical approach which may look
superficially similar to the one developed here.
For future reference, we discuss briefly the differences between
these two approaches.
De Haas et al use
\beq
\dot \gamma = {4\Sigma D\over 5 R \eta}\exp{64\pi\kappa\over 15 T} D^2 ,
\ee
in our notation. The corresponding expression within our 
approach derived from (\ref{eq:D})
reads
\beq
\dot \gamma = {4D \Sigma\over 5 \eta R}( 1+{6\kappa\over \Sigma R^2})
 \left(1-\left({11D\over 6}\right)^2\right)^{-1/2} 
\ee  in dimensionalized units. 
The $\kappa$ term in the first parenthesis arises from keeping bending energy
in the force balance . De Haas et al deliberately ignore bending
as being
irrelevant for large tension which  is true in the tense and
 upper part
of the entropic regime. Then  both expression yield the same linear
behavior at small $D$. The non-linear regime, however, is markedly different
with no exponential term
in our theory. Note also that, within our approach, $\Sigma$ still
carries an implicit $\dot \gamma$ dependence. 

The second type of reported behavior cannot so easily be reconciled with the
present theory which predicts a stationary tanktreading shape for all
values of shear rates and excess area. However, the theory allows for
a transient damped oscillatory approach to this stationary state.
In the prolate regime, the relaxation time
\beq
\tau_2\equiv \eta R^3/\kappa \Gamma_2E_2
\ee
 can become large for large
enough excess area because then formally $E_2 \to 0$ as 
$\Delta \to \infty$. 
For large enough shear rate, the product of oscillation
frequency  $\dot \gamma$ with the relaxation time $\tau_2$
can become large.  In this case,
 many oscillations can occur before the stationary state is reached.
 Thus, after turning on shear flow
or changing the shear rate one expects 
 transient flipping motion and oscillatory
behavior.  Still, the present
theory predicts finally a settling into a stationary
tanktreading state. 

\section {Discussion}

For a fluid membrane in an external hydrodynamic field, we derived a 
deterministic equation of motion taking into account both viscous dissipation 
in the
 surrounding liquid and local incompressibility of the membrane. For 
quasi-spherical vesicles in linear external flow, this equation decouples in
an expansion in spherical harmonics.
The area constraint
is implemented by an effective tension that depends on shear rate.
 Adding noise with correlations adapted from equilibrium, we obtain
a Langevin-type equation. Its solution for shear flow yields transient
 oscillations which settle into a stationary tanktreading state. The inclination
angle and the deformation parameter are both calculated as a 
function of shear rate and excess area. For small shear rate, the
inclination angle becomes $\pi/4$. With increasing shear rate it decreases
the stronger the larger the excess area is. 
The deformation parameter increases
linearly with
shear rate with a slope depending on excess area. 

This work is complementary to our previous study where we solved the
deterministic equations numerically for arbitrary 
excess area ignoring fluctuations \cite{z:krau96}.
 Taken together, the two approaches yield a fairly complete picture
for the behavior of fluid vesicles in all phase space. There remains only one
explored region. For an excess area too large for the quasi-spherical approximation
to apply and a shear rate too small for the numerics to handle, neither
approach is well-suited. In this regime, one would naively expect that small
shear leads to small distortions around the fluctuating equilibrium shape.
However, this remains to be checked by an explicit calculation.

A relatively straightforward extension of the present theory would be
the inclusion of further dissipative mechanisms such as 
inter-monolayer drag and shear within each monolayer. While both
types will certainly affect the transient behavior, it is less likely
that monolayer shear viscosity changes the stationary results for
quasi-spherical vesicles since the rotational component of the flow does
not lead to shear within the layer. Likewise, one could easily allow for
different viscosities inside and outside the vesicle.

The analytical progress possible within the
quasi-spherical approximation comes with two
 intrinsic weaknesses which cannot easily be circumvented. First,
in the prolate regime, this
approximation gives only a rough representation of the stationary
ellipsoidal mode since it does not produce a non-zero $u_{2,m}$ without
shear. Second, this approximation neglects higher order
couplings between deformation and flow field which can become 
relevant at larger shear rates. However, the relatively good
agreement between the present theory and the previous numerical work
shows that this effect is certainly not dominant for $v\gsim 0.8$.

The present study should encourage further experimental work. 
By measuring the inclination angle and the deformation parameter of a
single vesicle over the whole range of shear rates and fitting those
against theory using the full equation (\ref{eq:delta}),
  one should be able to extract fairly
precise data on excess area and bending rigidity especially when
combined with a variation of temperature \cite{z:haec95}. 

Concerning the experimentally reported oscillatory or flipping motion, it
seems to early for a final assessment. While it could be that the above
mentioned effects mask such a motion within the theory, there are two
more 
indications for doubting that 
flipping motion is genuine for non-spherical vesicles. 
First, our numerical work \cite {z:krau96}
 which did not suffer from the spherical approximation, 
did  not show flipping motion for at least $\chi \gsim 0.1$. Secondly, contrary to the reference given
by de Haas et al \cite{z:haas97}, previous work on tanktreading of red-blood-cells
\cite{kell82} 
did not predict flipping motion if the viscosities inside and outside are the 
same. Only if the inner viscosity is significantly larger that the outer,
a transition to flipping motion occurs.

\acknowledgments

Helpful discussions with R. Bruinsma, M. Kraus,
J. Prost, and H. Rehage are
gratefully acknowledged.

\appendix

\section{Adaption of Lamb's solution}
\def\v{{\bf v}}
\def \out{^{out}}
\def\in{^{in}}
Lamb's solution tells us the velocity field
$\v(\r)$ and pressure field $p(\r)$ in all space
if the velocity field is specified on a sphere with radius $R$.
We adapt the presentation of this classical problem given in Ref.
\cite{happ73} whose notation we follow. 
Rather than characterizing $\v(R,\theta,\phi)$ by its three components, Lamb's
solution uses the three quantities
\beq
X(\theta,\phi) \equiv {\bf V}(\r){\bf e}_r  ,
\ee
\beq
Y(\theta,\phi) = - R\ {\nabla }{\bf V}(\r)  ,
\ee
and
\beq
Z(\theta,\phi) = R \ {\bf e}_r({\nabla }\times {\bf V}(\r))  ,
\ee  
 derived from the field ${\bf V}(\r) \equiv {\bf v}(R,\theta,\phi)$.
Note that  ${\bf V}(\r)$ is independent of $r$.
Since the rotational component $Z$ cannot be excited by either
bending moments or  inhomogeneities in the surface tension, 
we can ignore it
in the following.

Given these boundary values of the velocity, the total velocity field
for $r<R$ becomes
\beq
{\bf v}\in(\r) = \sum_{l \geq 2}\left( \nabla \Phi_l\in(\r) + 
{l+3\over 2 \eta (l+1)(2l+3)} r^2 \nabla p_l\in(\r)
 - {l\over \eta (l+1)(2l+3)}r p_l\in(\r)\right)
\ee
where
\beq
\Phi_l\in(\r)= \sum _{m=-l}^{l} \Phi_{l,m}\in \Y (r/R)^l
\ee
and
\beq
p_l\in(\r) = \sum _{m=-l}^{l} p_{l,m}\in \Y (r/R)^l .
\ee
The expansion coefficients $ \Phi_{l,m}\in ,p_{l,m}\in$
are determined by the boundary conditions as
\beq
 \Phi_{l,m}\in(\r) = {R\over 2l}[(l+1)X_{l,m} - Y_{l,m}]
\ee
and
\beq
p_{l,m}\in(\r) =  {\eta(2l+3)\over lR}[Y_{l,m} - (l-1) X_{l,m}]\ee

For $r>R$, the velocity field can be obtained formally by
replacing $l$ by $-l-1$ in these expressions. Explicitly, it
reads 
\beq
{\bf v}\out(\r) = \sum_{l\geq 1}\left( \nabla \Phi_l\out(\r) + 
{-l+2\over 2 \eta (-l)(-2l+1)} r^2 \nabla p_l\out(\r)
 - {-l+1\over \eta (-l)(-2l+1)}r p_l\out(\r)\right)
\ee
where
\beq
\Phi_l\out(\r) = \sum _{m=-l}^{l} \Phi_{l,m}\out \Y (r/R)^{(-l+1)}
\ee
and
\beq
p_l\out(\r) = \sum _{m=-l}^{l} p_{l,m}\out \Y (r/R)^{(-l+1)} 
\ee
with
\beq
 \Phi_{l,m}\out(\r) = {R\over 2(l+1)}[l X_{l,m} + Y_{l,m}]
\ee
and
\beq
p_{l,m}\out(\r) =  {\eta(2l-1)\over (l+1) R}[Y_{l,m} + (l+2) X_{l,m}]  .
\ee

\def\T{{\bf T}}
\def\ns{{\bf }\nabla^S}

This velocity field leads to a stress vector acting across the surface of
a sphere of radius $r=R$ as
\beq
\T_R \equiv T_R {\bf e}_r + \T_t\equiv \left(
 -{\bf e}_r p + \eta \left({\p{\bf v} \over \p r} - 
{{\bf v}\over r}\right) + {\eta\over r}{\bf \nabla(rv)}\right)_{|r=R} .
\ee 
The normal component of this stress vector at $r=R$ reads
 for the inner solution
\beq
T_R\in= \sum_{l,m}\left(2(\eta/R^2) (l-1)l\Phi_{l,m}\in + 
(la_l\in - b_l\in)p_{l,m}\in\right) \Y  ,
\ee
where 
\beq
a_l\in\equiv {l(l+2)\over (l+1)(2l+3)}
\ee and
\beq
b_l\in\equiv {2l^2+4l+3\over (l+1)(2l+3)} .
\ee
Likewise, the 
normal component of the stress vector at $r=R$ reads
 for the outer solution
\beq
 T_R\out = \sum_{l,m} \left( [2(\eta/R^2) (l+2)(l+1)\Phi_{l,m}\out -[(l+1)a_{l}\out 
+ b_l\out)p\out_{l,m}]\right)\Y
\ee 
where
\beq
a_l\out\equiv {l^2-1\over l(2l-1)}
\ee 
and
\beq
b_l\out\equiv {2l^2+1\over l(2l-1)}  .
\ee
The tangential part at $r=R$ reads for the inner solution 
\beq
\T_t\in = 
\sum\left( 2 (\eta/R^2)(l-1) \Phi_{l,m}\in + a_l\in  p_{l,m} \in \right)\ns\Y ,
\ee 
and for the outer solution
\beq
\T_t\out = \sum\left( -2 (\eta/R^2)(l+2)  \Phi_{l,m}\out + a_l\out
p_{l,m}\out \right)\ns \Y  ,
\ee 
where 
\beq
\nabla^S\equiv{1\over R \sin\theta}{\bf e}_\phi \p_\phi + 
{1\over R}{\bf e}_\theta\p_\theta .
\ee

Accomodation of these well-known properties of Lamb's solution to our
problem requires to balance the stress discontinuities at the
vesicle surface with those exerted by the vesicle's 
bending moments and its inhomogeneous surface tension.
The key approximation involved for quasi-spherical vesicles is to
assume that all stresses act on a sphere of radius $R$ rather than on
the time-dependent vesicle's shape. This approximation can be controlled
for small external flow and small excess area. However, in the absence of
a workable alternative, we will use it for all
flow strengths. 
The stress balance  thus reads 
\beq
\T\out_{R} = \T\in_{R} + {1 \over \sqrt{g}}{\delta
F\over \delta \R}  .
\ee
The stress exerted by the vesicle follows from (\ref{eq:se})
in an expansion in
spherical harmonics as 
\beq
{1 \over \sqrt{g}}{\delta
F\over \delta \R} = (\kappa/R^2) \sum  
\left(E_l\u +2\sigma_{l,m})\Y{\bf e}_r -  \sigma_{l,m}
\ns\Y\right) .
\ee
Evaluating the tangential balance leads to
\begin{eqnarray}
(\kappa/\eta R^2) \sigma_{l,m} = & &\left({l-1\over l}\right)[(l+1)X_{l,m} - Y_{l,m}]
+ \left({l+2\over l+1}\right)  [l X_{l,m} + Y_{l,m}] \cr
& &+a_l\in \left({2l+3\over l}\right) [Y_{l,m} - (l-1) X_{l,m}]
-a_l\out\left({2l-1\over l+1}\right)[Y_{l,m} + (l+2) X_{l,m}] \cr
= & & {2l+1\over l(l+1)}(X_{l,m} + 2 Y_{l,m}).
\end{eqnarray}
Evaluating the normal balance, using this result for $\sigma_{l,m}$,
and setting $X_{l,m} =\Xi_{l,m}$ and $Y_{l,m}=\Yi_{l,m}$
leads to the relations (\ref{eq:Xi})-(\ref{eq:Bl}) quoted in the main part.

%\bibliographystyle{pr99sty}
%\bibliography{/usr/lan/teTeX/texmf/bibtex/bib/membranen,preprints,shear-foot}

\end{document}